\documentstyle[twocolumn,aps]{revtex}

\begin{document}
\draft
\title{Casimir energy for a scalar field with a frequency dependent boundary condition}

\author{H.\ Falomir, K.\ R\'{e}bora}
\address{IFLP - Dept.\ de F\'\i sica, Fac.\  de Ciencias Exactas,
Universidad Nacional de La Plata, C.C. 67, (1900) La Plata,
Argentina}
\author{and M.\ Loewe}
\address{Facultad de F\'\i sica, Pontificia Universidad Cat\'olica de Chile,
Casilla 306, Santiago 22, Chile}

\date{\today}

\maketitle

\begin{abstract}

We consider the vacuum energy for a scalar field subject to a
frequency dependent boundary condition. The effect of a frequency
cut-off is described in terms of an {\it incomplete}
$\zeta$-function. The use of the Debye asymptotic expansion for
Bessel functions allows to determine the dominant (volume, area,
\dots) terms in the Casimir energy. The possible interest of this
kind of models for dielectric media (and its application to
sonoluminescence) is also discussed.

\end{abstract}

\pacs{12.20.Ds, 03.70.+k, 78.60.Mq}


\section{Introduction}

The Casimir effect \cite{Casimir,Mostepanenko} arises as a
distortion of the vacuum energy of quantized fields due to the
presence of boundaries (or nontrivial topologies) in the
quantization domain. This effect, which has a quantum nature
associated with the zero-point oscillations in the vacuum state,
is significant in diverse areas of physics, from statistical
physics to elementary particle physics and cosmology.

In particular, in the last years there has been a great interest
in the Casimir energy of electromagnetic fields in the presence of
dielectric media, due to Schwinger's suggestion \cite{Schwinger}
that it could play a role in the explanation of the phenomenon of
sonoluminescence \cite{Putterman}.

The results obtained on this subject by different groups through
several calculation techniques (as Green's functions methods, van
der Waals forces, $\zeta$-function methods and asymptotic
developments for the density of states - see references
\cite{Milton-Ng-1,Milton-Ng-2,Brevik-Milton,Hoye-Brevik,Bordag-Kirsten-V,Lambiase,MP-V,Carlson-Visser,Liberati-Visser-1,Liberati-Visser-2,Nest-Pir}
among others) are rather controversial, and some basic issues
remain to be clarified. In this respect, it is our aim to
contribute to the understanding of the problem by studying a
simplified model, which incorporates a frequency cut-off in the
boundary conditions at the separation between media, to emulate
the behavior of real dielectrics.

It should be mentioned that authors introducing a cut-of in the
wave number to describe the behavior of real dielectrics agree
with Schwinger's explanation of the phenomenon (See, for example,
\cite{MP-V,Carlson-Visser,Liberati-Visser-1,Liberati-Visser-2}).
On the other hand in Ref.~\cite{Nest-Pir}, where the permittivity
of a non-magnetic medium is modeled by means of a (angular
momentum dependent) one-absorption-frequency Sellmeir relation,
and a $\zeta$-function technique is employed to sum-up the
contributions of the proper oscillation modes of the
electromagnetic field, it is concluded that the vacuum energy of
a bubble embedded in this material has the wrong behavior with
the radius (besides its absolute value being far too small) to be
relevant to sonoluminescence.

\bigskip

Candelas \cite{Candelas} has been the first to remark the
importance of dispersion in connection with the Casimir effect.
Later, Brevik and collaborators
\cite{Brevik1,Brevik2,Brevik3,Brevik4} have considered several
models with spherical and cylindrical geometries, showing that
dispersive effects may reverse the direction of the Casimir force
acting on the boundary between media. In
Ref.~\cite{Brevik1,Brevik2} it is studied the case of spherically
symmetric media satisfying the relation $\epsilon(\omega)
\mu(\omega)=1$ (of special interest in QCD), both with a sharp
frequency cut-off on $\mu(\omega)$ and with a behavior analogous
to the one-absorption-frequency Sellmeir's formula for
dielectrics. In Ref.~\cite{Brevik3}, for non-magnetic dispersive
media with spherical geometry, and based upon the Minkowski
energy-momentum tensor, the Casimir surface force is worked out
when the permittivity $\epsilon(\omega)$ presents a step along the
imaginary frequency axis. In the limit of perfect conductivity, a
non-dispersive term is recovered which is in agreement with Boyer
result \cite{Boyer,M-deR-S}, while an attractive dispersive
contribution is also found.

\bigskip

In what follows we consider a simple model of a scalar field
subject to frequency dependent local boundary conditions on the
surface of a sphere. Thus our main goal is to establish a method
for calculating the change of the Casimir energy of the field when
the radius of the sphere is varied, in a situation where the
boundary conditions impose a {\it physical} frequency cut-off
$\Omega$.

To this end we consider the very simple case of a scalar field
whose modes corresponding to eigenfrequencies $\omega \leq \Omega$
are confined to the interior of a sphere of radius $R$, satisfying
local homogeneous boundary conditions.

On the other hand, we will assume that the boundary is completely
transparent for those modes with $\omega > \Omega$. Therefore,
their contribution to the difference of Casimir energies for two
different values of $R$ will cancel out, no matter the
regularization employed for its definition. Consequently, we will
subtract these contributions, which amounts to a redefinition of
the reference energy level in an $R$-independent way.

For the evaluation of the vacuum energy of the low frequency
modes we will employ asymptotic expansions in an {\it
incomplete}-$\zeta$ summation technique, to be discussed in the
following. This approach will allow for the identification of the
{\it volume, surface, \dots,} dominant terms in the Casimir
energy.

As a final exercise we will show that, with reasonable values for
the frequency cut-off and the bubble radius, the amount of Casimir
energy available in this model is comparable with the energy
emitted during each cycle of a typical sonoluminesce experiment
\cite{Putterman}. In view of the controversy existing on the
subject, it would be of great interest to apply this method to a
similar model for the case of the electromagnetic field in the
presence of dielectric media, calculation which will be presented
in a forthcoming paper \cite{K-H}.

\section{The model and its incomplete $\zeta$-function}\label{partial-zeta}

Let us consider a free scalar field in $\mathbf{R}^3$ satisfying
at the surface of a sphere of radius $R$, local boundary
conditions which depend on the frequency $\omega$ of the field
modes.

We will make the assumption that the boundary is completely
transparent for the modes of frequencies grater than a cut-off
$\Omega$, while for $\omega \leq \Omega$ the modes satisfy
Dirichlet boundary conditions,
\begin{equation}\label{modos}
  \begin{array}{c}
    \left( \triangle + \frac{\omega^2}{c^2} \right) \psi_\omega(\vec{r})=0 ,
    \ {\rm for\ } r<R, \\ \\
   \left. \psi_\omega(\vec{r}) \right|_{r=R} = 0,
  \end{array}
\end{equation}
being confined to the interior of the sphere.

Writing $\psi_\omega(\vec{r})= f_l(r) Y_l^m(\theta,\varphi)$, we
get for the radial function
\begin{equation}\label{radial}
  \displaystyle{\left( \frac{d^2}{dr^2}+
  \frac{2}{r}  \frac{d}{dr} - \frac{l(l+1)}{r^2}
  + \frac{\omega^2}{c^2} \right)} f_l(r)=0, \ {\rm for} \ r<R,
\end{equation}
where the eigenfrequencies are determined by imposing the
condition
\begin{equation}\label{Dirich}
 \left. f_l(r) \right|_{r=R} = 0.
\end{equation}
The solutions regular at the origin are given by $f_l(r)=
\displaystyle{\sqrt{\frac{\pi}{2 z}}} J_\nu(z)$, with $\nu=l+1/2$
and $z=\omega_{\nu,n} r/c$, where the eigenfrequencies are
\begin{equation}\label{ceros}
  \omega_{\nu,n}  = \displaystyle{ \frac{c}{R} \, j_{\nu,n} },
\end{equation}
being $j_{\nu,n}$ the $n$-th zero of the Bessel function
$J_\nu(z)$.

\bigskip

We will be interested in the difference between the vacuum
energies of two situations differing in the value of $R$. Then,
we can disregard the contributions of those modes with frequencies
$\omega>\Omega$ because, being independent of the position of the
boundary, their contributions cancel out (whatever the
regularization employed in defining the vacuum energy would be).
This simply amounts to an $R$-independent subtraction, which is
nothing but a redefinition of the zero energy level.

Therefore, we should evaluate the (finite) sum
\begin{equation}\label{sum-cas}
  E(R) = \sum_{\nu=1/2}^{\nu_{0}} 2\,\nu \  \sum_{n=1}^{N_{\nu}} \
   \frac{1}{2} \,
  \hbar \, \omega_{\nu,n},
 \end{equation}
where $N_\nu$ is the number of positive zeroes of $J_\nu(z)$ less
than or equal to $x=\Omega R/c$, the factor $2\,\nu = 2\,l+1$ is
the eigenvalue degeneracy, and $\nu_{0}$ is the maximum value of
$\nu$ for which $N_\nu \geq 1$.

We are interested in an analytic, rather than numeric, evaluation
of eq.\ (\ref{sum-cas}). Although this is a finite sum, we will
employ a summation method based on the evaluation of an {\it
incomplete} $\zeta$-function, an approach which could be applied
in more complex situations. We can employ the following
representation:
\begin{equation}\label{sum-s}
  \sum_{n=1}^{N_{\nu}} \ j_{\nu,n} =
  \left. \sum_{n=1}^{N_{\nu}} \ j_{\nu,n}^{-s} \right|_{s=-1},
\end{equation}
where the sum in the right hand side obviously  exists for any
$s\in \mathbf{C}$ \footnote{Notice that the sum in the r.h.s.\ of
eq.\ (\ref{sum-s}) evaluated at $s=0$ gives $N_\nu$, the number of
eigenfrequencies contributing to the Casimir energy of the field
(after the subtraction made to define it) for a given value of
the angular momentum $l=\nu-1/2$.}.

Since $J_\nu(z)$, for $\nu > -1$, has only real zeros, and its
non-vanishing zeros are all simple \cite{G-R}, we can employ the
Cauchy theorem to represent the sum in the r.h.s.\ of
(\ref{sum-s}) as an integral on the complex plane,
\begin{equation}\label{integral}
  \sum_{n=1}^{N_{\nu}} \ j_{\nu,n}^{-s} =
  \displaystyle{\frac{1}{2\pi \imath}} \oint_C z^{-s}
  \displaystyle{\frac{J'_\nu(z)}{J_\nu(z)}} \  dz,
\end{equation}
where the curve $C$ encircles counterclockwise the $N_{\nu}$ first
positive zeros of $J_\nu(z)$.

For $\Re (s)$ large enough, the contour $C$ can be deformed into
two straight vertical lines, one crossing the horizontal axis at
$\Re (z)=x$ and the other at $\Re (z) = 0^+$. Indeed, expressing
the integrand in terms of the modified Bessel function\cite{A-S}
\begin{equation}\label{mod-Bes}
    I_\nu\left( w\right)=e^{{-\imath\frac{\pi}{2}\nu}}
    \,J_\nu\left(e^{{i \frac{\pi}{2}}}
    w\right)
\end{equation}
(valid for $-\pi < \arg(w)\leq \pi/2$), and taking into account
its asymptotic behavior for large arguments \cite{A-S}, it is
easily seen that, for $0<x \neq j_{\nu,n}, \forall n$, the
integral
\begin{equation}\label{zeta}
    \zeta_\nu(s,x)\equiv
  \displaystyle{\frac{-1}{2\pi \imath}}
  \int_{x-\imath\infty}^{x+\imath\infty} z^{-s}
  \displaystyle{\frac{J'_\nu(z)}{J_\nu(z)}} \  dz,
\end{equation}
converges absolutely and uniformly in the open half-line $s > 1$,
from which it can be meromorphically extended to the whole complex
$s$-plane.

Therefore, for $s>1$,
\begin{equation}\label{dif-zetas}
    \sum_{n=1}^{N_{\nu}} \ j_{\nu,n}^{-s} =
     \zeta_\nu(s,0^+) -  \zeta_\nu(s,x).
\end{equation}
And, since the left hand side of (\ref{dif-zetas}) is holomorphic
in $s$, the singularities of $ \zeta_\nu(s,x)$ must be independent
of $x$.

On the other hand, for $y>0$  \cite{A-S},
\begin{equation}\label{menos-z}\begin{array}{c}
   I_\nu\left( -y - i x\right)=
  e^{\displaystyle{-\imath\pi\nu}}
  I_\nu\left( y +  i x\right) \\ \\
  I_\nu\left( y +  i x\right)=
   \left( I_\nu\left( y -  i x\right) \right)^*.
\end{array}
\end{equation}
So, for real $s>1$ we can write
\[ \zeta_\nu(s,x)= \]
\begin{equation}\label{zeta-final}
  \Re \left\{ \frac{-x^{1-s}}{\pi}\,
  e^{\displaystyle{- i \frac{\pi}{2}(s+1)}}
  \int_0^\infty (y-\imath)^{-s}\,
  \displaystyle{\frac{I'_\nu\left(x( y - \imath) \right)}
  {I_\nu\left(x( y - \imath) \right)}}\, dy \right\} .
\end{equation}

In order to construct the analytic extension of $\zeta_\nu(s,x)$
to $s\simeq -1$, we subtract and add to the integrand in
(\ref{zeta-final}) the first few terms obtained from the uniform
asymptotic (Debye) expansion \cite{A-S} of the Bessel functions,
\begin{equation}\label{Debye-exp}
   \displaystyle{
    \frac{I'_\nu\left(\nu\, t \right)}{I_\nu\left(\nu\, t \right)}
     = \frac{1}{\nu} D_{\nu}(t)
     + {\mathcal O}(\nu^{-3})},
\end{equation}
where
\begin{equation}\label{Debye-expDir}\begin{array}{c}
   \displaystyle{ D_{\nu}(t)= \nu D^{(1)}(t) + D^{(0)}(t) +
    \nu^{-1} D^{(-1)}(t)= }\\ \\
    \displaystyle{\frac{{\nu\,\sqrt{1 + {t^2}}}}{t}} -
  \displaystyle{\frac{t}{2\,\left( 1 + {t^2} \right) } \,+\,}
     \displaystyle{\frac{4\,t - {t^3}}
    {8\,{\nu}\,
      {{\left( 1 + {t^2} \right) }^{{\frac{5}{2}}}}}}.
\end{array}
\end{equation}
valid for large $\nu$ with fixed $t$. We will see that this
approximation allows for the identification of the volume,
surface, \dots \ contributions to the vacuum energy.

Changing the integration variable in eq.\ (\ref{zeta-final}) to
$t\equiv z(y-\imath)$, with $z=x/\nu >0$, we get
\[ \zeta_\nu(s,x)= \]
\begin{equation}\label{zeta-final1}
  \Re \left\{ \frac{-\nu^{-s}}{\pi}\,
  e^{\displaystyle{- i \frac{\pi}{2}(s+1)}}
  \int_{- i z}^{\infty- i z} t^{-s}\,
  \displaystyle{\frac{d\left(\ln{I_\nu\left(\nu\, t\right)}\right)}
  {d\, t}}\, dt \right\} .
\end{equation}

So, we must consider the integral
\begin{equation}\label{sustrac}\begin{array}{c}
    \displaystyle{
  \int_{- i z}^{\infty- i z} t^{-s}\,
  \displaystyle{\frac{d\left(\ln{I_\nu\left(\nu\, t\right)}\right)}
  {d\, t}}\, dt=\displaystyle{
   \int_{- i z}^{\infty- i z} t^{-s}\, D_{\nu}(t)\, dt \   +
  }
  }\\ \\
   \displaystyle{ +
   \int_{- i z}^{\infty- i z} t^{-s}\,
  \left\{
  \displaystyle{\frac{d\left(\ln{I_\nu\left(\nu\, t\right)}\right)}
  {d\, t}} - D_{\nu}(t) \right\}\, dt
  }.
\end{array}
\end{equation}
The second integral in the right hand side of this equation
converges for $s>-2$, since the integrand can be estimated by
means of the next (${\cal O}(\nu^{-3})$) term in the Debye
expansion (eq.\ (\ref{Debye-exp})), which behaves as ${\cal
O}(t^{-3})$ for large $|t|$. It can be numerically evaluated at
$s=-1$. This will not be done in this paper.

\bigskip

In the following, we will consider only the first integral in the
right hand side of (\ref{sustrac}), retaining the first few terms
of its expansion in powers of $\nu^{-1}$ consistent with the
approximation made in eq.\ (\ref{Debye-exp}).

Notice that the integrand is an algebraic function, having
singularities at $t=0,\pm \imath$, and behaving as ${\cal O}(t^0)$
for large $|t|$. This integral converges absolutely and uniformly
for $s > 1$, where it defines an analytic function to be
meromophically extended to the region of interest of the
parameter $s$. As we will see, this extension reveals the
singularities of $\zeta(s,x)$ as simple poles, whose residues are
independent of $x$ (a necessary condition to give a finite result
in (\ref{dif-zetas}) for any $s$).

In fact, by virtue of the analyticity of the integrand, we can
change the path of integration to write
\[ \displaystyle{
   \int_{- i z}^{\infty- i z } t^{-s}\, D_{\nu}(t) \, dt }
   = \]
\begin{equation}\label{orden-dominante-appendix}\begin{array}{c}
   \displaystyle{
   \int_{- i z}^{1 } t^{-s}\,
    \left(  \displaystyle{\frac{\nu\,{\sqrt{1 + {t^2}}}}{t}} -
  \displaystyle{\frac{t}{2\,\left( 1 + {t^2} \right) } \, +}
     \displaystyle{\frac{4\,t - {t^3}}
    {8\,{\nu}\,
      {{\left( 1 + {t^2} \right) }^{{\frac{5}{2}}}}}} \right)  \, dt
      }
   \\ \\
  \displaystyle{ +
   \int_{1}^{\infty} t^{-s}\,
     \, \left\{ \displaystyle{\frac{\nu\,{\sqrt{1 + {t^2}}}}{t}} -
  \displaystyle{\frac{t}{2\,\left( 1 + {t^2} \right) }\, + }
     \displaystyle{\frac{4\,t - {t^3}}
    {8\,{\nu}\,
      {{\left( 1 + {t^2} \right) }^{{\frac{5}{2}}}}}} - \right. } \\ \\
         \displaystyle{ \left.
     - \,
        \left[\nu\,\left( 1 + \frac{1}{2\,t^2} \right) - \frac{1}{2\,t}
    -  \frac{1}{8\,\nu\,t^2}  \right] \right\} \, dt  +}  \\ \\
     \displaystyle{ +
   \int_{1}^{\infty } t^{-s}\,
  \left[\nu\,\left( 1 + \frac{1}{2\,t^2} \right) - \frac{1}{2\,t}
    -  \frac{1}{8\,\nu\,t^2}   \right] \, dt   } .
\end{array}
\end{equation}

The first integral in the r.h.s.\  of eq.\
(\ref{orden-dominante-appendix}), containing the whole dependence
on $x=\nu\,z$, is holomorphic in $s$ and can be directly evaluated
at the required value of this parameter. On the half-line
$(1,\infty)$ we have subtracted and added the first terms in the
series expansion of $D_{\nu}(t)$ for large $t$, thus making the
second integral to converge for $s>-2$. The third one must be
evaluated for $s>1$ and then analytically continued to the
relevant values of $s$. This  can be exactly done, its
contribution to $\zeta_\nu(s,x)$ in eq.\ (\ref{zeta-final1})
being the real part of
\begin{equation}\label{sing-orden-nu}
\begin{array}{c}
\displaystyle{  \frac{
 e^{\displaystyle{-\frac{ i }{2}\,\pi \,\left( 1 + s \right) } }}
 {{\nu}^{1 + s}\,8\,
        \pi }}\,
      \left( \frac{8\,{\nu}^2}{1 - s} + \frac{4\,\nu}{s} +
        \frac{1 - 4\,{\nu}^2}{1 + s} \right)
\end{array}
\end{equation}
This expression has simple poles at $s=0,\pm 1$, which are the
only singularities of $\zeta_\nu(s,x)$ for $\Re(s)>-2$. Notice
that the residues of $\zeta_\nu(s,x)$ are independent of $x$,
\begin{equation}\label{residuos}
  \begin{array}{c}
        \displaystyle{Res\,  \zeta_\nu(s,x)|_{s=1} = \frac{1}{\pi}\,  ,}\\ \\
        \displaystyle{Res\,  \zeta_\nu(s,x)|_{s=0} = 0\,  ,}\\ \\
        \displaystyle{Res\,  \zeta_\nu(s,x)|_{s=-1} =
        \frac{1 - 4\,{\nu}^2}{8\,\pi }\,  ,}
\end{array}
\end{equation}
and in agreement with the results in  \cite{Romeo} (where
$\zeta_\nu(s,0^+)$ is studied).

For example, for $\zeta_\nu(s,x)$ around $s=-1$ and for $\nu<x$
(which will be needed in Section \ref{dominantes} to evaluate the
vacuum energy), one straightforwardly obtains the Laurent
expansion
\begin{equation}\label{zeta-en-1}\begin{array}{c}
  \zeta_\nu(s,x>\nu)= \displaystyle{{\frac{1- 4\,{{\nu}^2}}
   {8\,\pi \,\left( 1 + s \right) }}+}\\ \\
  \displaystyle{+ \left[ \frac{
    \left( 4\,{\nu}^2 -1 \right) }{8\,\pi }\,
    \left[\log \left(\frac{\nu}{2}\right)+ \log \left(z+ \sqrt{z^2 -1}\right) \right]-
  \frac{\nu^2}{4\,\pi } + \right.} \\ \\ \displaystyle{\left.
  -  \frac{\nu^2}{2\,\pi } \, z\, \sqrt{z^2 -1} -
  \frac{3 z - 8 z^3}{24\,\pi\,\left( z^2 -1 \right)^{3/2}} - \frac{1}{3
  \pi}+ {\cal O}(\nu^{-1})\right] +} \\ \\ \displaystyle{+ {\cal O}(s+1)},
\end{array}
\end{equation}
with fixed $z=x/\nu\gtrsim 1$.

On the other hand, for $x\rightarrow 0^+$ a similar calculation
leads to
\begin{equation}\label{zeta-en-1-x=0}\begin{array}{c}
  \zeta_\nu(s,x=0^+)= \displaystyle{{\frac{1- 4\,{{\nu}^2}}
   {8\,\pi \,\left( 1 + s \right) }}
   + \left[
   {\frac{\left(  4\,{{\nu }^2-1} \right) }{8\,\pi }}\,
      \log \left({\frac{\nu }{2}}\right) - \right.}\\ \\ \displaystyle{\left.-
      {\frac{{{\nu }^2}}{4\,\pi }} -
  {\frac{\nu }{4}} - {\frac{1}{3\,\pi }} + {\cal
  O}(\nu^{-1})\right]+
        {\cal O}(s+1)}.
\end{array}
\end{equation}

In the following Section we will evaluate, as a function of $\nu$,
the number of modes contributing in eq.\ (\ref{dif-zetas}), and
in Section \ref{dominantes}, their contributions to the vacuum
energy.

\section{The number of contributing modes} \label{numero-de-modos}

In this Section we address ourselves to the determination of
$\nu_{0}$ in (\ref{sum-cas}), the maximum value of $\nu$ for
which $N_\nu \geq 1$. Although in the simple case under study the
zeros of $J_\nu(w)$ are well known \cite{A-S}, we prefer to
establish a criterium which can be applied in more general
situations.

First, notice that
\begin{equation}\label{dif-s0}
  N_{\nu}(x)\equiv \left.
  \sum_{n=1}^{N_{\nu}} \ j_{\nu,n}^{-s}\right|_{s=0}
  = \left. \left[\zeta_\nu(s,0^+) -
  \zeta_\nu(s,x)\right]\right|_{s=0}
\end{equation}
is a discontinuous function of $x$, having a step of height $1$ at
each positive zero $j_{\nu,n}$ of the Bessel function $J_\nu (w)$.

Then, $\nu_{0}$ can be determined from the condition
\begin{equation}\label{nu0}
  N_{\nu_0}(x)= N_{\nu_0}(j_{\nu_0,1}+0) =1,
\end{equation}
with $N_{\nu_0}(j_{\nu_0,1}-0) = 0$.

Taking into account eq.\ (\ref{residuos}) and the fact that the
second and third integrals in the r.h.s.\ of eq.\
(\ref{orden-dominante-appendix}) are real, it is straightforward
to obtain from eqs.\
(\ref{zeta-final1}-\ref{orden-dominante-appendix}) that
\begin{equation}\label{en-0}\begin{array}{c}
   \zeta_\nu(s=0,x) =
   \displaystyle{{-{\frac{\nu}{2}}- \frac{1}{4}}\,+ }
    \\ \\
    \displaystyle{
  +\Re \left\{ -
  {\frac{\imath\,\nu\,
      }{\pi }}\left({\sqrt{1 + {e^{-\imath\,  \pi }}\,{z^2}}}-
      \log(1 + {\sqrt{1 + {e^{-\imath\,  \pi
  }}\,{z^2}}})\right)
   \right. } \\ \\
  \displaystyle{ \left.
   +{\frac{\imath}{4\,\pi}}
  \log (1 +{e^{-\imath\,\pi}}\, {z^2}) +
  \right. } \\ \\ \displaystyle{ \left.
    +{\frac{\imath(2+3 z^2)}{24\,\nu\,\pi \,
      {{\left( 1 +{e^{-\imath\,\pi}}{z^2}
      \right) }^{3/2}}}}
   \right\} + {\cal O}(\nu^{-2})},
\end{array}
\end{equation}
where we have taken $z=x/\nu \approx 1$. In particular, for
$x\rightarrow 0^+$,
\begin{equation}\label{enx0}
  \zeta_\nu(s=0,x=0^+) =\displaystyle{{ {-\frac{\nu}{2}}-\frac{1}{4}} +
  {\cal O}(\nu^{-2})},
\end{equation}
in coincidence with \cite{Romeo}.

Now, taking the difference between (\ref{enx0}) and (\ref{en-0})
we get a {\it smooth} approximation, $\tilde{N}_\nu(x)+{\cal
O}(\nu^{-1})$, to the step function $N_\nu(x)$ in (\ref{dif-s0}).
It is easily seen that, for $\nu>x$, $\tilde{N}_\nu(x)=0$ while,
for $\nu<x$, we have
\begin{equation}\label{parazmay1}\begin{array}{c}
   \tilde{N}_\nu(x)= \displaystyle{ {\frac{\nu\,
      }{\pi }}\left( {\sqrt{{z^2}-1}}
     -\arctan( {\sqrt{{z^2}-1}}) \right) -
    } \\ \\
      \displaystyle{ -{\frac{1}{4}}
   -{\frac{2+ 3 z^2}{24\,\nu\,\pi \,
      {{\left( {z^2} -1
      \right) }^{3/2}}}} } ,
\end{array}
\end{equation}
with $z=x/\nu$.

Let us now determine the value $\nu_0$ for which
\begin{equation}\label{1/2}
  \tilde{N}_{\nu_0}(x)=1/2 .
\end{equation}
To this end, we propose an expansion of the form
\begin{equation}\label{des-mu}
  \sqrt{z_0^2 -1}=\varepsilon_1 \, \nu_0^{-1/3} +
  \varepsilon_3 \, \nu_0^{-3/3} +
  {\cal O}(\nu_0^{-5/3}),
\end{equation}
which makes sense for $\nu_0\gg 1$ and $z_0=x/\nu_0\approx 1$.
Replacing in (\ref{parazmay1}) and imposing (\ref{1/2}), the
coefficients $\varepsilon_k$ can be determined order by order in
$\nu_0^{-1/3}$, to get
\begin{equation}\label{xdenu0}
  x= \nu_0 + 1.857 \, \nu_0^{1/3} + 1.034 \, \nu_0^{-1/3} +
  {\cal O}(\nu_0^{-1})
\end{equation}
or, inverting this development,
\begin{equation}\label{nu0dex}
  \nu_0=x-1.857 \, x^{1/3} + 0.1155 \, x^{-1/3}+
  {\cal O}(x^{-1}).
\end{equation}
(Notice that $\nu_0<x$.)

Equation (\ref{xdenu0}) is in excellent agreement with the
expression of the first non-vanishing zero of $J_{\nu_0}(w)$, for
a large order $\nu_0$ \cite{A-S}: $
  j_{\nu_0,1}= \nu_0 + 1.8557 \, \nu_0^{1/3} + 1.03315 \, \nu_0^{-1/3} +
  {\cal O}(\nu_0^{-1})$.

\section{The dominant contributions to the vacuum energy}\label{dominantes}

In this Section we evaluate the first contributions to the vacuum
energy obtained from the Debye expansion employed in Section
\ref{partial-zeta}. As we will see, this allows to determine the
volume, surface, \dots , terms in the Casimir energy of the
scalar field.

According to the results in the previous Section, we are
interested in the Laurent expansion of $\zeta_\nu(s,x>\nu)$ and
$\zeta_\nu(s,x=0^+)$  around $s=-1$, given in eqs.\
(\ref{zeta-en-1}) and (\ref{zeta-en-1-x=0}). As already remarked,
the singular parts cancel out in the difference on the r.h.s.\ of
eq.\ (\ref{dif-zetas}) (see eq.\ (\ref{residuos})). For the
difference of the finite parts we get
\begin{equation}\label{diferencia}\begin{array}{c}
\displaystyle{ z\, F(x,\nu) \equiv
 \left[ \zeta_\nu(s,0^+) -  \zeta_\nu(s,x)\right]_{s=-1}=}\\ \\
  \displaystyle{
  = {\frac{{{\nu }^2} }{2\,\pi }} \,
      \left( z\,{\sqrt{-1 + {z^2}}} -
        \log (z + {\sqrt{-1 + {z^2}}}) \right) - }\\ \\
        \displaystyle{
   - {\frac{\nu}{4}} + {\frac{3\,z - 8\,{z^3}}
    {24\,\pi \,{{\left( -1 + {z^2} \right) }^
        {{\frac{3}{2}}}}}} +
  {\frac{1}{8\,\pi }}\, {\log (z +
       {\sqrt{-1 + {z^2}}})}
       }\\ \\ \displaystyle{+ {\cal O}\left(\nu^{-1}\right)}.
\end{array}
\end{equation}
This is a good approximation as long as $\nu \gg 1$ and $z=x/\nu
\gtrsim 1$.

\bigskip

Our aim is now to evaluate the sum in eq.\ (\ref{sum-cas}),
\begin{equation}\label{sum-cas-ap}\begin{array}{c}
 \displaystyle{
  E(R=x\, c/\Omega) = } \\ \\ \displaystyle{
  = \frac{\hbar\,\Omega}{x}\,
  \sum_{\nu=1/2}^{\nu_{0}} \,\nu \,
  \left[\zeta_\nu(-1,x) -  \zeta_\nu(-1,0^+)\right]=}  \\ \\
  \displaystyle{
  = \hbar\,\Omega \, \sum_{\nu=1/2}^{\nu_{0}} \, F(x,\nu),
  }
\end{array}
 \end{equation}
with $\nu_0$ given in (\ref{nu0dex}).

The function $F(x,\nu)$ is non-negative and has a pronounced
maximum at $\nu\approx x/2$ (i.e. $z\approx 2$). Thus, the use of
the approximation in eq.\ (\ref{diferencia}) is consistent if $x
\gg 1$.

From eq.\ (\ref{diferencia}), it is not difficult to see that the
successive terms in the Euler - Maclaurin summation formula
\cite{A-S},
\begin{equation}\label{Euler}\begin{array}{c}
  \displaystyle{
   \sum_{\nu=1/2}^{\nu_k} \, F(x,\nu)=
   \int_{1/2}^{\nu_0}\, F(x,\nu)\, d\nu +} \\ \\
   \displaystyle{ +  \frac{1}{2}\left\{
   F(x,\nu_0)+ F(x,1/2)
  \right\} +
   } \\ \\
  \displaystyle{
    +\left.  \frac{1}{12} \,
  \partial_\nu \left[F(x,\nu )\right]
  \right|^{\nu=\nu_0}_{\nu=1/2}+
  \dots
  }\, ,
  \end{array}
\end{equation}
are of increasing order in $x^{-1}$. So, retaining the first few
terms consistent with the approximation made, we get for the
vacuum energy
\begin{equation}\label{Euler1}\begin{array}{c}
  \displaystyle{
   \frac{E(R)}{\hbar\,\Omega}= }\\ \\
   x^3\,\left( \frac{-\left( {\sqrt{-1 + z_0^2}}\,
          \left( -5 + 2\,z_0^2 \right)  \right) }{24\,\pi \,
        z_0^3} + \frac{2\,z_0^4 -
        3\,\log (z_0 + {\sqrt{-1 + z_0^2}})}{24\,\pi \,z_0^4}
     \right)  - \\ \\
     - x^2\,\left( \frac{-{\sqrt{-1 + z_0^2}}}
      {4\,\pi \,z_0^2} + \frac{\pi  +
        3\,\log (z_0 + {\sqrt{-1 + z_0^2}})}{12\,\pi \,z_0^3}
     \right)  -\\ \\ -
      x\,\left( \frac{-\left( -27 + 17\,z_0^2
          \right) }{48\,\pi \,z_0\,{\sqrt{-1 + z_0^2}}} +
     \frac{6\,\pi  - 10\,z_0^2 +
        69\,\log (z_0 + {\sqrt{-1 + z_0^2}})}{48\,\pi \,z_0^2}
     \right)
   \\ \\
    + {\cal O}(x^0),
\end{array}
\end{equation}
where $z_0=x/\nu_0$, $x=R\, \Omega/c$ and $\nu_0$ is given in eq.\
(\ref{nu0dex}). In this equation one can recognize volume,
surface and curvature contributions to $E(R)$. Notice that we
could have retained any number of terms of the asymptotic
expansion in eq.\ (\ref{Debye-exp}), and performed the same steps
as in the present calculation, to get the Casimir energy to any
order in $x^{-1}$.

Finally replacing in eq.\ (\ref{Euler1}) $z_0\rightarrow x/\nu_0$
and $\nu_0$ by its expression we get
\begin{equation}\label{exp-final}
  \displaystyle{
  E(R)=
  }
  \displaystyle{ \hbar\,\Omega
  \left[
   {\frac{{x^3}}{12\,\pi }}-
  {\frac{{x^2}}{12}}-
  0.1343\,{x^{{\frac{4}{3}}}}+
   {\cal O}(x)
   \right] },
\end{equation}
where one can see that the volume and surface terms are dominant
for $x>>1$. Notice that also non-integer powers of the radius $R$
appear as a consequence of the relation between $\nu_0$ and $x$,
eq.\ (\ref{nu0dex}).

\section{Conclusions}

In equations (\ref{Euler1}-\ref{exp-final}) we have derived the
dominant contributions to the vacuum energy of a scalar field in
a model with a frequency dependent boundary condition, consisting
in the confinement of the modes with low frequency (up to a
physical cut-off $\Omega$) to the interior of a sphere of radius
$R$.

These modes are subject to Dirichlet boundary conditions at the
surface of the sphere, while those with frequency higher than
$\Omega$ are free, being the boundary completely transparent to
them. This characteristic of the model allows for the subtraction
of the contribution of the high frequency modes to the vacuum
energy, which amounts (independently of the regularization
employed to define it) to an $R$-independent redefinition of the
zero energy level, having no consequences on the evaluation of
energy differences.

In so doing, we have represented the sum over the eigenfrequencies
up to the cut-off $\Omega$ in terms of an {\it incomplete}
$\zeta$-function associated with the Laplacian operator in the
sphere with Dirichlet boundary conditions (see eq.\
(\ref{dif-zetas})). The function $\zeta(s,x)$ as in eq.\
(\ref{zeta-final}) is well defined only for $\Re(s)>1$. So, it
was analytically continued from $s>1$ to the relevant values of
this parameter ($s=0$, needed to evaluate the maximum angular
momentum $l_0 = \nu_0 -1/2$ giving rise to eigenfrequencies less
than or equal to $\Omega$, and $s=-1$, necessary to evaluate the
contribution to the Casimir energy of the modes with angular
momentum $l=\nu -1/2$) by approximating the behavior of the
integrand in eq.\ (\ref{zeta-final}) employing the Debye
asymptotic expansion of the modified Bessel functions appearing
in its expression.

This procedure has lead to a meromorphic function having simple
poles with (exactly evaluated) cut-off independent residues (see
eq.\ (\ref{residuos})), a necessary condition to have a finite
result for the sum in eq.\ (\ref{dif-zetas}).

The finite part of $\zeta(s,x)$ has been evaluated up to terms of
a given order in $\nu^{-1}$. Although we have retained only the
first terms in this asymptotic expansion, one can follow the same
steps to determine the finite part of $\zeta(s,x)$ at $s=-1$ or
$0$ up to any required precision in $\nu^{-1}$.

Finally, the application of the Euler - Maclaurin summation
formula has lead to an expression for the Casimir energy of the
model in which one can recognize volume, surface and curvature
contributions (see eq.\ (\ref{Euler1})).

For a cut-off corresponding to $x=R\, \Omega/c >> 1$, the
dominant terms in the vacuum energy, eq.\ (\ref{exp-final}), are
proportional to the volume ($V = 4 \pi R^3/3$) and area ($S=4 \pi
R^2$) of the sphere,
\begin{equation}\label{contrib-D}\begin{array}{c}
  \displaystyle{
    \frac{ E(R)}{ \hbar\,\Omega} =  V\, \frac{{\Omega }^3}{16\,\pi^2\,c^3}
    + \xi\, S\, \frac{{\Omega }^2}{12\,\pi^2\,c^2} + \dots\ }
\end{array},
\end{equation}
with $\xi=-\pi/4$.

It is worthwile to remark that, for a similar model where the low
frequency modes of the scalar field are subject to Neumann (rather
than Dirichlet) boundary condition, we get the same expression
for the dominant terms with $\xi=+\pi/4$.

These two dominant terms are in complete agreement with those
obtained from the expansion of the density of states in powers of
the inverse wavelength\footnote{Indeed, for scalar fields subject
to local homogeneous boundary conditions the density of states is
modified by finite volume effects \cite{Balian1}. The first
correction in the asymptotic expansion for large wavelength $k$
is given by
\begin{equation}\label{dens-estados}\begin{array}{c}
  \displaystyle{
  \sum_{n} \sim   V\, \int\,\frac{d^{3}k}{(2\,\pi)^{3}}\,
    + \, S\, \int\,\xi\,\frac{d^{3}k}{(2\,\pi)^{3}\,k} + \dots}
\end{array},
\end{equation}
where the coefficient $\xi$ takes the value $\xi=- \pi/4$, $\xi=+
\pi/4$ and $\xi=+ \pi/4$ for Dirichlet, Neumann and Robin
boundary conditions respectively. Then, introducing the dispersion
relation $\omega(k)= c\,k$ and a cut-off in the wavelength given
by $K=\Omega/c$, it is easy to get eq.\ (\ref{contrib-D}) for the
vacuum energy.

For {\it junction} boundary conditions, as those appearing in
problems with dielectrics, the coefficient $\xi$ has more involved
expressions, which can be difficult to evaluate \cite{MP-V}.}
\cite{Balian1,MP-V}. The relation between $\nu_0$ and $x$, eq.\
(\ref{nu0dex}) (or, equivalently, the expression of the first
zero of the Bessel function $J_{\nu_0}(w)$ in terms of the order
$\nu_0$ \cite{A-S}) introduces also non-integer powers of the
radius $R$ (see eq.\ (\ref{exp-final})).

\bigskip

As a final exercise, we can use eq.\ (\ref{exp-final}) in a very
schematic model pretending to mimic the phenomenon of
sonoluminescence. We will adopt the values of the radius and
emitted energy corresponding to a typical sonoluminescent bubble
\cite{Putterman}, and estimate the cut-off $\Omega$ needed to
produce this amount of energy. To this end, we will simply take
the difference of the low frequency contribution to the Casimir
energy of the scalar field for two different values of the bubble
radius.

If the bubble collapses from an initial radius $R=4\times
10^{-5}\, m$ to a final radius of one tenth this value, and the
emitted energy is ${\cal E}=1.2\times 10^{-12}\, Joule$, by
imposing the equality
\begin{equation}\label{cut-off}
  \displaystyle{
   {\frac{R}{\hbar\,c}}\,\left(E(R)- E(R/10)\right) =
   1.516\times{{10}^9},
    }
\end{equation}
and taking into account eq.\ (\ref{exp-final}), it follows that
$x=490$, justifying the use of the approximation obtained.

This implies that $\Omega=3.675\times{{10}^{15}}\ 1/sec$, which
corresponds to a cut-off in wavelengths in the ultraviolet of
$\Lambda=5.129\times{{10}^{-7}}\ m = 5129 \ \AA$, not far from
the region where the refraction index of water becomes
essentially 1 \cite{Jackson}. This strongly suggests to consider a
similar model for the case of the electromagnetic field in the
presence of dielectric media, calculation which will be presented
elsewhere \cite{K-H}.

\bigskip

{\bf Acknowledgements:} The authors thank E.M.\ Santangelo for
useful comments. This work was partially supported by Fundaciones
Andes - Antorchas under Contract Nr.\ C-13398/6, ANPCyT
(Argentina) - PICT'97 Nr.\ 00039, CONICET (Argentina) - PIP Nr.\
0459, and FONDECyT (Chile) - grants Nr.\ 1980577 and 7980011.






%
%

%
%

\end{document}